\documentclass[useAMS,usenatbib]{mn2e}
\usepackage[pdftex]{graphicx}

\newcommand{\htwo}{${\rm H_2}$}
\newcommand{\hsix}{${\rm H_6^+}$}
\newcommand{\hd}{${\rm (HD)_3^+}$}

\title[Snowflakes in heaven]{A snowflake's chance in heaven}
\author[Walker]{Mark A. Walker\thanks{Mark.Walker@manlyastrophysics.org}\\
$\!\!$Manly Astrophysics, 3/22 Cliff St,  Manly 2095, Australia\\}
\begin{document}

\date{Accepted ................................ Received ................................ In original form ................................}

\pagerange{\pageref{firstpage}--\pageref{lastpage}} \pubyear{2013}

\maketitle

\label{firstpage}

\begin{abstract}
We consider the survival of solid \htwo\ in the diffuse interstellar medium, with application to grains which are small enough to qualify as dust. Consideration of only the thermal aspects of this problem leads to the familiar conclusion that such grains sublimate rapidly. Here we show that charging plays a critical role in determining the sublimation rate, because an electric field helps to bind molecules to the grain surface. A key aspect of the charging process is that the conduction band of solid hydrogen lies above the vacuum free-electron energy level, so low-energy electrons cannot penetrate the solid. But they are attracted by the dielectric and by positive ions in the matrix, so they become trapped in vacuum states just above the surface. This charge-separated configuration suppresses recombination and permits overall neutrality, while supporting large electric fields at the surface. Charging  ceases when the potential energy just outside the electron layer coincides with the conduction band minimum. By that stage the heat of sublimation has increased tenfold, effecting a huge reduction in the sublimation rate. Consequently hydrogen grains may be able to survive indefinitely in the diffuse ISM. There are good prospects for identifying \htwo\ grains, if they exist, as fully-localised surface electrons should exhibit discrete energy levels, with a corresponding spectral line signature.
\end{abstract}


\section{Introduction}
Solid hydrogen is thought to be absent from the diffuse interstellar medium (ISM). The reason is simple: typical pressures in the diffuse medium are far below the saturated vapour pressure of \htwo\ at all accessible temperatures, so sublimation is very rapid (Greenberg and de~Jong 1969; Field 1969). This appears to be a compelling argument for the non-existence of interstellar solid hydrogen because it only relies on knowledge of the phase-diagram of \htwo, and that has been accurately determined in the laboratory. It is, however, misleading. Laboratory experiments exclude the various ionising-  and charged-particles which pervade interstellar space and, as this paper demonstrates, that makes a big difference. When surface charging is considered, it appears that hydrogen grains may be able to survive indefinitely in the diffuse ISM.

This paper is tightly focused on the thermodynamics of \htwo\ surfaces in an interstellar context. As our starting point we take macroscopic grains of pure hydrogen, in contrast to circumstances where H$_2$ molecules form a monolayer around other species. Hydrogen monolayers may be important both at the nanoparticle level (Duley 1996; Bernstein, Clark and Lynch 2013), and as coatings on other bulk materials (Sandford and Allamandola 1993).  But the electronic properties of these composites differ from those of bulk, solid H$_2$ and we do not consider them here. 

Even if we had no reason to suspect the presence of snowflakes, the overwhelming abundance of hydrogen in astronomical contexts would still make it worth communicating the results reported herein. But to motivate readers to engage  with this material it behoves us to add some context around the narrow issue of snowflake survival, as follows. Hydrogen snow cannot be formed in the diffuse ISM --- it requires gas in self-gravitating clouds. The clouds must be sufficiently cold and dense that they lie close to the saturated vapour-pressure curve of \htwo\ (see figure 1 of this paper). Such clouds would be difficult to detect -- i.e. they would be a form of baryonic dark matter -- and current observational constraints are not able to exclude a Galactic population of this type. Various specific models have been proposed, including a Galactic thin-disk population of massive, fractal clouds (Pfenniger, Combes and Martinet 1994; Pfenniger and Combes 1994), a disk population of spherical clouds (Lawrence 2001), and a Galactic halo population of spherical clouds (e.g. Henriksen and Widrow 1995; Gerhard and Silk 1996; Walker and Wardle 1998; Draine 1998; Sciama 2000; Walker 2007). The likely presence of solid \htwo\ within such clouds was pointed out by Pfenniger and Combes (1994), and it was argued by Wardle and Walker (1999) that snowflakes can confer thermal stability on the self-gravitating gas.

There are two ways in which such clouds can seed the diffuse ISM with hydrogen dust. First, as they move through the Galaxy, small amounts of material -- including \htwo\ grains -- may be stripped from their surfaces and left behind in the diffuse ISM. Secondly, an individual cloud may be heated faster than it can cool by radiation.  In this circumstance the cloud must cool by expanding, causing precipitation of a substantial fraction of the \htwo. If the total heat supplied suffices to unbind the cloud, this solid material will be injected into the diffuse ISM. Examples of such disruptive heating events include the UV flash from a nearby supernova explosion, and a low-speed (a few ${\rm km\,s^{-1}}$) collision between a pair of clouds. We note that the cloud-cloud collision rate may be significant over the life of a galaxy (Gerhard and Silk 1996; Walker 1999).

The other important contextual point relates to the spectral features expected from particles of solid hydrogen. Longward of the far-UV,  \htwo\  has no electric-dipole transitions and the pure solid consequently displays only very weak absorption. Inevitably, though, an initially pure \htwo\ grain would acquire trace molecular ion species as a result of ionising particles impinging on it, and the electric-dipole transitions of these ions provide a signature of the solid. It is known that the ionisation chemistry of solid \htwo\ differs greatly from the gas phase: whereas H$_3^+$ is the dominant ionisation product in \htwo\ gas, in the condensed phase \hsix\ is favoured (Lin, Gilbert and Walker 2011, and references therein).  This molecular ion was discovered only recently and its infrared spectrum has yet to be measured in the lab. But it is a small molecule that has been tackled with a high level of ab initio electronic structure theory, yielding accurate wavenumbers for five fundamental vibrational transitions (Lin, Gilbert and Walker 2011).  And for the two isotopomers which are expected to dominate in practice (i.e. \hsix\ and \hd), these authors found a striking match to the observed pattern of mid-infrared spectral bands of the ISM. What those spectral coincidences mean is at present unclear, but to sort out that issue we certainly need a good understanding of the durability of interstellar solid hydrogen. This paper is a step towards that goal.

It was pointed out by Lepp (2003, citing Dyson and Williams 1997), in the context of pre-galactic chemistry, that free charges enhance the stability of hydrogen grains, because of the energy associated with polarisation of the \htwo\ molecules. Here we explore that idea further for the case of grains in the diffuse ISM of our Galaxy. We restrict attention to collisional charging, which occurs because of the dilute plasma in which the grains are immersed. Although photoelectric charging of solid \htwo\ is potentially  important, there has been no experimental investigation of the photoelectric yield and so the charging rates are currently unknown, even to order of magnitude. We do know that the band-gap of solid \htwo\ (approximately 14.5~eV: Inoue, Kanzaki and Suga 1979) is greater than the ionisation energy of atomic hydrogen. For grains immersed in the typical interstellar radiation field, that means that the rate of photoexcitations into the conduction band is many orders of magnitude smaller than for graphite or silicate grains (Bakes and Tielens 1994; Weingartner and Draine 2001). It is also small compared to the rate of photodissociations, which we identify in \S2.2 as the main source of grain heating.

In the case of graphite or silicates the dynamics of collisional charging limit the magnitude of the surface voltage to be of order the thermal energy in the plasma (Spitzer 1941). For that reason it might be imagined that the surface fields of charged hydrogen dust are strongly dependent on grain size. That turns out not to be the case. Hydrogen grains develop a surface charge distribution which is akin to a plasma double-layer, with the outer layer shielding the field of the inner one. Consequently the most important length-scale is the layer separation of a few \AA ngstr\"oms, with grain size playing a subordinate role.

The structure of this paper is as follows. In the next section we recap phase equilibrium, sublimation rates and lifetimes for uncharged \htwo\ grains. Section 3 describes the properties of vacuum electronic states above the surface of solid \htwo. We then turn to collisional grain charging in \S4. In \S5 we revisit \htwo\ phase equilibrium and grain sublimation rates, this time accounting for the influence of macroscopic electric fields on the heat of sublimation. These calculations demonstrate that interstellar hydrogen grains may be long-lived.

\section{Thermodynamics of pure hydrogen}
Survival of \htwo\ grains in the ISM has been considered by Greenberg and de~Jong (1969) and by Field (1969). E.S.~Phinney (1985, unpublished manuscript) gave a thorough treatment of the sublimation of \htwo\ snowballs in a cosmological context. The treatment given here follows Phinney's development, but adapted to small particles in the diffuse ISM of our own Galaxy.

\subsection{Phase equilibrium of \htwo}
Suppose we have a gas of \htwo\ at temperature $T$ and pressure $P$, in thermodynamic equilibrium with a lump of solid \htwo. The flux of molecules, each of mass $m$, incident upon the solid is
\begin{equation}
F_{in}={P\over\sqrt{2\pi m kT}},
\end{equation}
and over the temperature range of interest to us here almost all of the incident molecules are expected to stick to the surface. In thermodynamic equilibrium the pressure of \htwo\ must equal the saturated vapour pressure, $P_{sat}$:
\begin{equation}
P_{sat}=kT{{(2\pi m kT)^{3/2}}\over{h^3}}\exp\left(-{{b_o}\over{kT}}-\Delta\right),
\end{equation}
where $h$ is Planck's Constant, $b_o$ is the molecular heat of sublimation, and $\Delta\equiv\pi^4(T/\Theta)^3/5$, with $\Theta\simeq105\,$K the Debye temperature of the solid. The value of $b_o$ can be obtained by comparing equation (2) with measurements. For {\it para-\/}\htwo\ at the Triple Point ($T_{tp}=13.8\;$K), Souers (1986) gives\footnote{The data refer to \htwo\ whose nuclear spin statistics equilibrated at $T=20.4$~K, so the {\it ortho-\/}\htwo\ content is small, but not zero.} $P_{sat}=7,\!030\,$Pa, from which we deduce $b_o/k=90.7\;$K.

In equilibrium there is no net flux of molecules from the surface, so the sublimation rate, $F_{out}$, must equal the flux arriving, $F_{in}$. We thus obtain
\begin{equation}
F_{out}=kT{{2\pi m kT}\over{h^3}}\exp\left(-{{b_o}\over{kT}}-\Delta\right),
\end{equation}
as the sublimation rate at temperature $T$.

The kinetic energy flux of the molecules arriving at the surface of the grain is $2kTF_{in}$. On average, each arriving molecule releases an amount of heat equal to the heat of sublimation, $b_o$, so the grain heating rate per unit area due to the arriving molecules is $(b_o+2kT)F_{in}$. In equilibrium this must equal the grain cooling rate due to departing molecules, and the sublimation cooling rate per unit area is  therefore\footnote{Phinney (1985) gives $b_o+5kT/2$ for the mean energy carried away per sublimating molecule.}
  $(b_o+2kT)F_{out}$.

\begin{figure}
\includegraphics[width=85mm]{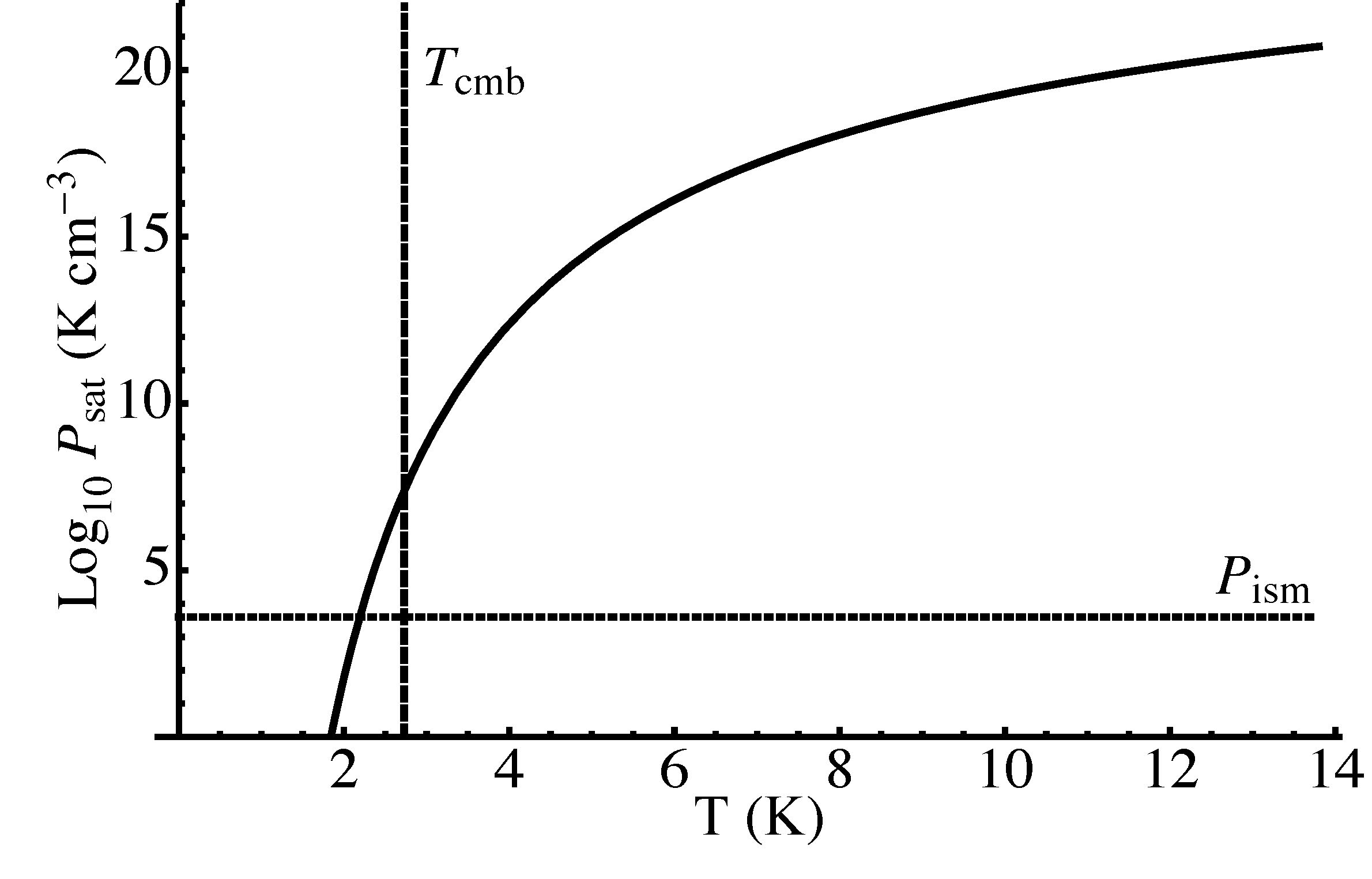}
\vskip-0.4truecm
\caption{The saturated vapour pressure of \htwo\ (solid curve) for temperatures below the triple-point ($T_{tp}=13.8\,$K). The vertical dashed line marks $T=T_{cmb}=2.73\,$K, and the horizontal dashed line shows the typical pressure in the diffuse ISM, $P_{ism}\sim4,\!000\,{\rm K\,cm^{-3}}$. The fact that $P_{sat}\gg P_{ism}$ for all $T\ge T_{cmb}$ argues that phase equilibrium between solid and gaseous phases of pure \htwo\ cannot be achieved in the diffuse ISM.}
\medskip
\end{figure}

Because of the exponential factors in equations (2) and (3), both $P_{sat}$ and $F_{out}$ are strong functions of temperature. The form of $P_{sat}(T)$ is shown in figure 1, from which we see that $P_{sat}$ greatly exceeds typical pressures in the diffuse ISM ($P_{ism}\sim4,\!000\;{\rm K\,cm^{-3}}$; Jenkins and Tripp 2011), even for $T=T_{cmb}=2.73\,$K, the temperature of the Cosmic Microwave Background (CMB). As the steady-state temperature of the ISM must exceed $T_{cmb}$, this graph leads us to expect that solid \htwo\ cannot be in phase equilibrium with the diffuse interstellar gas. Consequently if the diffuse ISM contains any hydrogen grains we expect them to be shrinking by sublimation.

\subsection{Thermal equilibrium of \htwo\ grains}
Equations (1) and (3) correctly describe the influx and efflux of molecules even if the phases are not in equilibrium. So for $P\sim P_{ism}\ll P_{sat}$ we can neglect $F_{in}$ in comparison with $F_{out}$ and treat the sublimation as if it is occuring in vacuum. The sublimation rate depends on the grain temperature, and that in turn is set by the balance between heating and cooling. Sublimation itself contributes to cooling, as does thermal photon emission from the grain. On the other side of the equation there is heating due to absorption from the InterStellar Radiation Field (ISRF), including the CMB, and due to impinging material particles, i.e. cosmic-rays and the thermal particles which constitute the interstellar gas.  

Material particles have significant energy-densities in the ISM, and an interaction cross-section that is close to the geometric cross-section of the grain. However, grains are highly transparent to cosmic-rays, with each particle depositing only a tiny fraction of its energy in the grain, so cosmic-ray heating is negligible. Thermal particles deposit all of their energy in the grain, but they move so slowly that their contribution to grain heating is also small compared to the absorption of starlight.

An unusual aspect of solid \htwo\ is that energy absorbed from a radiation field is not necessarily thermalised in the crystal as a whole, but may remain localised as electronic and/or ro-vibrational excitations on the absorbing centre. It is for this reason that solid \htwo\ is used for matrix isolation spectroscopy (e.g. Oka 1993; Anderson et al 2002). Therefore in determining the grain temperature we can neglect heating due to the ISRF except where there is a manifest coupling to lattice vibrations. This is the case for microwave radiation, which couples directly to the phonon modes of the crystal, and for far-UV radiation which can dump mechanical energy into the lattice as a result of dissociation of the \htwo\ molecules. Henceforth we use the term ``far-UV'' to refer specifically to wavelengths $\lambda$ such that $912<\lambda({\rm\AA})\la 1100$, i.e. the band of the ISRF which contributes the bulk of the dissociating photons. More energetic photons, such as X-rays, are negligible by comparison, because of the combined effects of low intensity in the typical ISRF and high transparency of micron-sized \htwo\ grains.

We adopt the analytic description of the ISRF given by Draine (2011), based on the tabulation by Mathis, Mezger and Panagia (1983). For that description the far-UV energy-density is $U_{fuv}=9.6\times10^{-15}\,{\rm erg\,cm^{-3}}$, with a mean photon energy of $\langle h\nu\rangle=12.2\,$eV. The rate of far-UV energy absorption by a grain of cross-section $\sigma$ is $U_{fuv}\sigma\, c$. In this paper we will consider only spherical grains of radius $a\ga0.1\,{\rm\mu m}$, for which the geometric cross-section, $\sigma=\pi a^2$, is an adequate approximation in the far-UV band. (Such grains also experience only small temperature fluctuations, $\la0.01\,$K, on absorption of a single far-UV photon.) All \htwo\ molecules in the solid are expected to be in the ro-vibrational ground state, for which the absorption of a far-UV photon by an isolated \htwo\ molecule leads to a dissociation probability of $p_{dis}\simeq0.13$ (Draine and Bertoldi 1996). We assume that the same average branching ratio holds for solid \htwo. Subtracting the energy required for dissociation ($G_0=4.48\,$eV; Herzberg 1969), we infer a lattice heating rate due to dissociations of
\begin{eqnarray}
\dot{E}_{dis}=p_{dis}U_{fuv}\left(1-{{G_0}\over{\langle h\nu \rangle}}\right)\pi a^2 c\quad\;\cr
\simeq 7.6\times10^{-13}\left(\!{a\over{1\,{\rm\mu m}}}\right)^{\!2}\;{\rm erg\,s^{-1}}.
\end{eqnarray}

Microwave heating/cooling depends on the low-frequency behaviour of the dielectric constant of solid \htwo, $\epsilon=\epsilon_1+i\epsilon_2$. The real part was measured by Constable, Clark and Gaines (1975) to be $\epsilon_1=1.2833$, with very little temperature variation across the range $0.3 - 3.3\,$K.  Insulators are expected to follow $\epsilon_2\propto1/\lambda$, at low frequencies, but the constant of proportionality is uncertain for solid hydrogen. Here we use the same assumption as Wardle and Walker (1999),  namely $\epsilon_2=\lambda_2/\lambda$, with $\lambda_2=1\,{\rm\mu m}$.  For $\epsilon_2\propto1/\lambda$ and $a\ll1\,$mm, the net microwave cooling rate is proportional to $T^6-T_{cmb}^6$. For our adopted value of $\lambda_2$, balancing photodissociation heating plus net microwave heating with sublimation cooling  leads to an equilibrium grain temperature of $T_{eq}=2.42\,$K, for a grain of radius $a=1\,{\rm\mu m}$. This is not a true steady state as the grain is shrinking and will eventually vanish.

With the grain temperature below that of the CMB, increasing the microwave coupling (larger $\lambda_2$, or larger $a$) leads to greater heating by the CMB and therefore slightly greater equilibrium temperatures. However, the dependence of $T_{eq}$ on $a$ and $\lambda_2$ is very weak because sublimation cooling and photodissociation heating are the dominant processes, and sublimation cooling is very temperature sensitive.

Later in this paper we will be examining circumstances where the sublimation rate is drastically lowered by surface charging of \htwo\ grains. In that case the sublimation cooling is  negligible and thermal equilibrium is achieved by a balance between photodissociation heating and net microwave cooling. With our choice of $\lambda_2$, the equilibrium temperature for a grain of $1\,{\rm\mu m}$ radius would then be $7.4\,$K (less for larger grains). That is substantially cooler than radiative equilibrium temperatures estimated for micron-sized silicate grains in the diffuse ISM (Draine 2011). Hydrogen grains are expected to be cool because they thermalise only a tiny fraction ($\sim10^{-3}$) of the starlight in the ISRF.\footnote{If our adopted microwave dielectric constant is extrapolated to shorter wavelengths it implies that the matrix will absorb far-IR radiation from the ISRF. Including that as an additional heat source for the matrix raises the equilibrium temperature of a micron-sized grain from $7.4\,$K to $7.8\,$K.}

Previous authors have considered composite grain structures in which a solid \htwo\ mantle surrounds a core of graphite (e.g. Wickramasinghe and Reddish 1968; Hoyle, Wickramasinghe and Reddish 1968; Wickramasinghe and Krishna Swamy 1969). In that case the graphite core would thermalise a substantial fraction of the ISRF,  so a composite grain is expected to be warmer than  a pure \htwo\ grain.

\subsection{Lifetime of$\,$ \htwo\ $\!\!$ grains}
The sublimation timescale of an interstellar hydrogen grain of radius $a$ is
\begin{equation}
t_{sub}=a {{\rho_*}\over{mF_{out}}},
\end{equation}
where $\rho_*=0.087\,{\rm g\,cm^{-3}}$ is the density of solid \htwo. For $T=2.42\;$K this evaluates to $t_{sub}\simeq6\times10^9(a/1\,{\rm\mu m})\,$s. In the absence of a mechanism for supplying dust to the ISM on a similar, or shorter timescale, this calculation suggests that the abundance of interstellar \htwo\ grains ought to be negligible. 

Any \htwo\ grain which has a temperature different from $T_{eq}=2.42\;$K will be driven towards that equilibrium value by the imbalance between heating and cooling. The temperature evolution is governed by
\begin{equation}
C{{{\rm d}T}\over{{\rm d}t}}=\dot{E},
\end{equation}
where the heat capacity of the grain is (Debye model)
\begin{equation}
C=16\pi a^3 {{\rho_*}\over{m}} k \Delta,
\end{equation}
and $\dot{E}$ is the net heating rate. A fiducial value is $\dot{E}=\dot{E}_{dis}$
for which we can estimate the corresponding heating timescale $t_{heat}\sim10^5(a/1\,{\rm\mu m})\,$s at $T=T_{eq}=2.42\,$K. The heating timescale is small compared to the sublimation timescale, so the lifetime of an \htwo\ grain is not lengthened significantly if it is initially very cold.

\smallskip
The foregoing calculations appear to preclude any possibility of \htwo\ grains persisting in the diffuse interstellar medium. However, it is clear that the sublimation rate and the saturated vapour pressure are both very sensitive to the heat of sublimation, $b_o$, so any effect which significantly increases the binding of \htwo\ molecules to the surface can affect this conclusion.

\smallskip
As noted earlier, the polarisation of \htwo\ molecules associated with static electric fields increases their heat of sublimation. The rest of this paper is devoted to grain charging and the influence of charges on the longevity of \htwo\ grains. Before modelling the charging process itself, we first describe the nature of the electronic surface states above solid \htwo. 

\section{External surface-state electrons}
There is an extensive literature on Surface State Electrons (SSE) above liquid helium, solid hydrogen and, to a lesser extent, solid neon (e.g. Cole 1974; Grimes 1978; Edel'man and Faley 1983; Leiderer 1992; Smolyaninov 2001). Bound surface states arise when there is a surface barrier which prevents electrons from penetrating the solid. Such a barrier exists for condensed noble gases, and for \htwo, because at an atomic/molecular level these species have closed electron shells which strongly repel any additional electrons. Consequently the electronic structure of these materials exhibits a conduction band which lies above the vacuum electron energy level.

For solid \htwo\ the barrier height calculated using the Wigner-Seitz method is $V_0=3.27\,$eV (Cole 1970). An early experimental investigation of field emission in liquid  \htwo\ suggested a low barrier ($0.3\pm0.2\,$eV: Halpern and Gomer 1969), but subsequent experiments with liquid \htwo\ have yielded results which are in close accord with predictions based on the Wigner-Seitz model (Johnson and Onn 1978). We therefore adopt Cole's (1970) theoretical value of the barrier height.

Whereas the surface barrier prevents electrons from entering solid \htwo, the electrical polarisation which they induce nevertheless attracts them to the dielectric. Any positive ions lodged in the matrix likewise attract electrons to the surface, and as a result there exists an electron potential well just outside the solid. That well supports bound electron states. Low energy electrons which are incident on the surface cannot penetrate the solid, but in the process of reflection from the surface they excite lattice vibrations, lose kinetic energy and may become trapped in bound vacuum states. Once there is a significant SSE population, incident electrons may also lose energy by scattering from that population, which increases the capture efficiency (i.e. the fraction of incident electrons which are captured into bound, vacuum states). For the purposes of our charging calculations (\S4) we assume that the capture efficiency is unity if the incident electron has insufficient energy to penetrate the solid.

\subsection{Image-charge bound states}
Laboratory investigations of the properties of SSE have all been undertaken in the limit of low charge densities and small applied fields. For this circumstance the attraction of an electron to the surface can be described by the vacuum potential associated with the image-charge (i.e. dielectric polarisation). In this section we summarise relevant results from the review of Cole (1974).

An electron near a planar dielectric boundary has potential energy
\begin{equation}
-e\Phi(z)=-{{\xi e^2}\over{4\pi\epsilon_o z}},\qquad(z>0),
\end{equation}
due to interaction with its image charge. Here $z$ is the distance above the surface, and
\begin{equation}
\xi\equiv{{\epsilon_1-1}\over{4(\epsilon_1+1)}}.
\end{equation}
For solid \htwo\ we have $\epsilon_1=1.2833$ (Constable, Clark and Gaines 1975), hence $\xi=0.0310$. The Schr\"odinger equation separates into a trivial, free-particle problem in the plane of the surface, and confinement in the perpendicular dimension. Given the potential in equation (8) it is straightforward to obtain analytic solutions in the limit where the surface barrier is treated as infinite. In this limit the energy eigenvalues are
\begin{equation}
E_n=-\xi^2{{\cal R}\over{n^2}}+{{p_\perp^2}\over{2m_e}},
\end{equation}
where ${\cal R}$ is the Rydberg Constant, $n$ is a positive integer, and $p_\perp$ is the electron momentum in the plane of the surface. The values of $p_\perp$ are not quantised, so electrons bound in these states can be thought of as a two-dimensional, free-electron gas. For solid \htwo\ the lowest energy level is $E_1=-13\,$meV, which is sufficiently small in magnitude (i.e. $|E_1|\ll V_0$) that an infinite barrier is a good approximation.

The corresponding wavefunctions are also ``hydrogenic'' in $z$, being products of the associated Laguerre polynomials with an exponential factor. The ground state wavefunction is
\begin{equation}
\psi_1={{2z}\over{\sqrt{z_o^3}}}  \exp[-z/z_o], 
\end{equation}
where $z_o=1/\xi$ in atomic units. For solid \htwo\ $z_o\simeq17.1\,$\AA. Evidently the approximation of a planar surface is acceptable for the grain sizes we are considering ($a\ga0.1\,{\rm\mu m}\gg z_o$). Surface electron states around small dielectric particles have been considered by Rosenblit and Jortner (1994). (See also Khaikin, 1978, who proposed a possible connection with interstellar grains.)

The solutions given in equations (10) and (11) are appropriate to the early stages of grain charging. However, in this paper we are particularly interested in the late stages, where the confining potential is primarily due to a population of sub-surface positive ions. We now turn to that case.

\subsection{Uniform-field bound states}
In the presence of a high surface-density of positive ions, bound in the matrix of the solid, the potential immediately above the surface can be approximated by
\begin{equation}
\Phi(z)=\Phi(0)-{\cal E}z,\qquad(z>0),
\end{equation}
where ${\cal E}$ is the electric field. We again assume a large surface barrier for electrons, approximating this with the limit $-e\Phi(z<0)\rightarrow\infty$. The wavefunctions appropriate to this potential are Airy Integrals:
\begin{equation}
\psi_k(z)={1\over{\sqrt{s}{\rm Ai}^\prime(-\zeta_k)}} {\rm Ai}\left({z\over{s}}-\zeta_k\right),
\end{equation}
with energy eigenvalues 
\begin{equation}
E_k=-e\Phi(0)+\zeta_k e\,{\cal E}s+{{p_\perp^2}\over{2m_e}},
\end{equation}
where
\begin{equation}
s\equiv\left[{{\hbar^2}\over{2 m_e e {\cal E}}}\right]^{1/3},
\end{equation}
and ${\rm Ai}(-\zeta_k)=0$. In our subsequent development we restrict attention to the ground state, $k=1$, for which the appropriate eigenvalue is $\zeta_1\simeq2.338$.

\subsection{Exterior potential profile of a hydrogen grain}
We will see in \S4 that charging tends to lead to a grain which is overall neutral. Thus, far from the grain surface the SSE shields out the electric field of the sub-surface ions, whereas there is no shielding at all at the grain surface. To correctly model the fluxes of charged particles reaching the grain surface, we need to describe the resulting potential variation. 

Because the electron layer is so thin, $z_o, s\ll a$, the potential variation through the SSE can be approximated by a step function. The amplitude of the step, $\delta\Phi_{sse}$, can be determined by integrating Poisson's equation through the SSE, with the volume charge density given by $-eN_e|\psi|^2$, where $-eN_e$ is the surface charge density in the SSE. If the charge density in the solid matrix is small, the appropriate wavefunction is that given in equation (11), whereas equation (13) (with $k=1$) is a better approximation for large values of $N_p$. In either case the result can be expressed as
\begin{equation}
\delta\Phi_{sse}  =  \mu {{e}\over{ \epsilon_o}}N_e,
\end{equation}
with the appropriate length scale being $\mu=1.5z_o$ for the image-charge confining potential (exact result), and $\mu\simeq1.56s$ for a linear potential (approximate result obtained from numerical integration). The characteristic scale $s$, in the wavefunctions of equation (13), decreases as the surface charge density increases. Those wavefunctions are appropriate to high surface charge densities. Thus the potential step through the SSE can be approximated by equation (16), with
\begin{equation}
\mu = {\rm min}(1.5 z_o, 1.56s).
\end{equation}
We adopt this prescription in our model of hydrogen grain charging in \S4.

Exterior to the SSE layer our model potential is simply Coulomb. If the surface density of positive charges embedded in the solid is $N_p$ (which may be positive or negative),  then we have
\begin{equation}
\Phi_{c}(r)  =  {{e a^2}\over{ \epsilon_o r}} (N_p-N_e),
\end{equation}
at radii such that $r-a\gg\mu$. 

\section{Collisional charging}
The rate at which charged particles impinge upon the surface of an uncharged grain can be written in the same form as equation (1). In the case of a charged grain, the flux of charged particles is increased or decreased, depending on the sign of the charge, because of the associated attraction or repulsion. The magnitude of this effect depends on $\gamma=-V/kT_{ism}$, the ratio of electrostatic potential energy at the grain surface to the characteristic thermal energy in the gas. Henceforth we assume for simplicity that in the interstellar gas the positive ions are protons and that they have the same number density as the electrons. Denoting the ionisation fraction by $x_e$, the resulting fluxes are (Spitzer 1941):
\begin{equation}
F_{e,p}={P_{ism}\over\sqrt{2\pi kT_{ism} m_{e,p}}}\,\left({{x_e}\over{1+x_e}}\right)\,{\cal F}(\gamma_{e,p}),
\end{equation}
where 
\begin{eqnarray}
{\cal F}(\gamma)&=1+\gamma,\qquad\quad\;(\gamma\ge0)\\
&\;=\exp(\gamma),\qquad\;\;\,(\gamma<0).
\end{eqnarray}

The net number-density of positive charges lodged in the solid matrix, $N_p$, evolves according to
\begin{equation}
{{{\rm d}N_p}\over{{\rm d}t}}=F_p(\gamma_p)-F_e(\gamma_e),
\end{equation}
i.e. the flux of protons, minus that of electrons. Defining ${\cal N}_p\equiv N_p/N_0$, with $N_0:= kT_{ism}\epsilon_o/e^2a$, and $\tau\equiv t/t_0$, with $t_0:=N_0/F_e(0)$, this becomes
\begin{equation}
{{{\rm d}{\cal N}_p}\over{{\rm d}\tau}}=\left({{m_e}\over{m_p}}\right)^{\!\!1/2}{\cal F}(\gamma_p)-{\cal F}(\gamma_e).
\end{equation}
The relationships between the two $\gamma$'s and the net density of ions on the surface differ markedly between solid \htwo\  and more conventional grain materials. The conventional materials are simpler in this respect and we briefly review that case before turning to solid hydrogen.

\begin{figure}
\includegraphics[width=85mm]{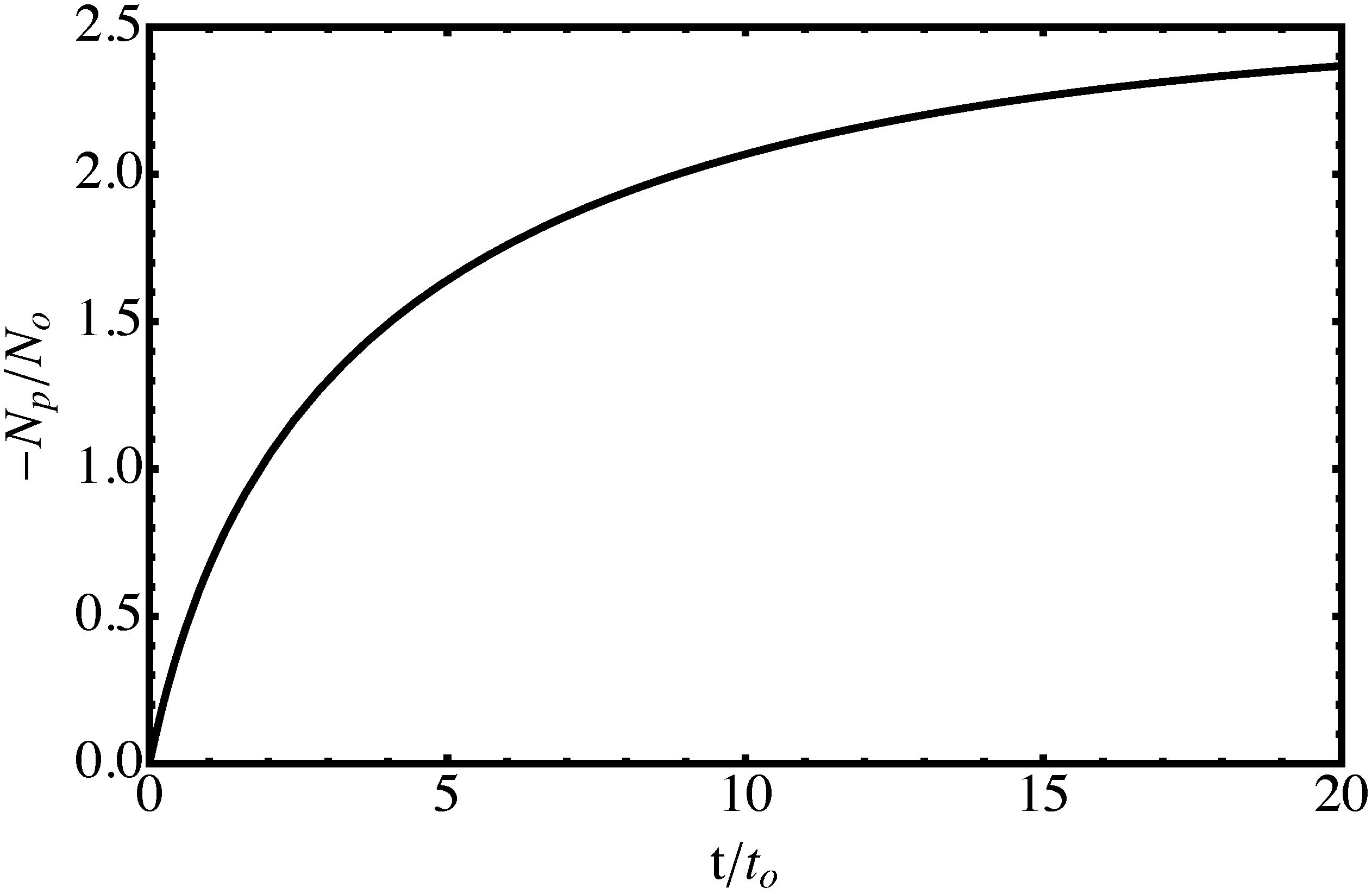}
\caption{The development of net charge on a grain of conventional composition (i.e. silicate or graphite), in the case of pure collisional charging. This graph shows the solution to the differential equation (23), starting from an uncharged grain at $t=0$.}
\medskip
\end{figure}

\subsection{Conventional grain materials}
For silicates and graphitic grains, any electrons or ions which reach the grain surface will stick there and then recombine at some rate. Neglecting the polarisation of the grain, which can be important in some circumstances (Draine and Sutin 1987), the external potential is Coulomb and is completely determined by the net charge on the grain. In this case we have
\begin{equation}
\gamma_e={\cal N}_p=-\gamma_p,
\end{equation}
and we can proceed to solve our differential equation (23), subject to the initial condition ${\cal N}_p=0$ at $\tau=0$. 

Initially the derivative in equation (23) is approximately equal to $-1$ and a net negative charge accumulates on the grain with ${\cal N}_p\simeq-\tau$. Subsequently the net negative charge means that the electron (proton) flux decreases (increases), so the rate of charging diminishes. At late times the rate of charging asymptotically approaches zero, and ${\cal N}_p\rightarrow-{\cal N}_\infty$, with
\begin{equation}
(1+{\cal N}_\infty)\exp({\cal N}_\infty)=\sqrt{m_p/m_e}.
\end{equation} 
The solution to equation (25) is the familiar result for collisional grain charging\footnote{If, contrary to our assumption, the positive ions in the diffuse ISM include a significant fraction of metals (which are more massive than protons), then the right-hand-side of equation (25) may increase by up to a factor of a few, with the result that ${\cal N}_\infty$ may increase by a few tens of percent.} (Spitzer 1941) ${\cal N}_\infty\simeq2.504$. Figure 2 shows the evolution of ${\cal N}_p$ with $\tau$, and table 1 gives the characteristic values of $N_0$ and $t_0$ for a grain of $a=1\,{\rm\mu m}$ radius in various phases of the ISM. In table 1, and throughout the rest of this paper, we adopt the usual acronyms for ISM phases: CNM, Cold Neutral Medium; WNM, Warm Neutral Medium; WIM, Warm Ionised Medium; and HIM, Hot Ionised Medium.

\begin{table}
\caption{Collisional charging characteristics in four ISM phases}
\begin{tabular}{@{}lccccr}
\hline
Phase  & $T_{ism}$ & $x_e$ &$F_e(0)$                          & $N_0$                   &$t_0$ \\
             &(K)                &             & ${\rm(cm^{-2}\,s^{-1})}$ & ${\rm(cm^{-2})}$ & ${\rm(s)}$\\
\hline
CNM &   $10^2$                 &   $2\times10^{-4}$    &   $10^{4}$               &   $5\times10^{7}\,\ $  &   $4\times10^{3}$  \\
WNM &   $5\times10^{3}$  &   $2\times10^{-2}$   &   $2\times10^{5}$  &   $2\times10^{9}\,\ $  &   $10^{4}$  \\
WIM &   $10^{4}$                 &   $1$                           &   $3\times10^{6}$  &   $5\times10^{9}\,\ $  &   $10^{3}$  \\
HIM &   $10^{6}$                  &   $1$                           &   $3\times10^{5}$  &   $5\times10^{11}$  &   $10^{6}$  \\
\end{tabular}

\medskip
{Conventional abbreviations are used for the phases of the ISM, viz: CNM, Cold Neutral Medium; WNM, Warm Neutral Medium; WIM, Warm Ionised Medium; HIM, Hot Ionised Medium. The numerical values of $N_0$ and $t_0$ quoted here are appropriate to grains of radius $a=1\,{\rm\mu m}$; these quantities both scale $\propto1/a$. Estimates of the ionised fraction, $x_e$, are taken from Draine (2011).}
\end{table}

\subsection{Solid \htwo\ grains}
The collisional charging of solid hydrogen grains differs from the behaviour just described. First, in order to enter the solid, electrons must reach the surface with sufficient energy to overcome the barrier, $V_0$, presented by the band structure of the lattice. Secondly, those electrons which have insufficient energy to penetrate the lattice will become trapped in the potential well just above the surface. We therefore have to keep track of the surface density of this two-dimensional electron gas, $N_e$, in addition to the net density of positive charges on the grain itself.  Finally, the presence of electrons above the surface causes the external potential to deviate from the Coulomb form; we approximate that deviation by the potential step $\delta\Phi_{sse}$ given in equations (16), (17).

We can compute the particle fluxes for a general potential as follows. Suppose the radial profile of the potential energy is $V(r)$ for a particular species, and that $V\rightarrow0$ for $r\rightarrow\infty$. If the energy-at-infinity is $E_\infty$, then a general expression describing the cross-section for the particle to reach radius $a$ is
\begin{equation}
\sigma(a)={\rm min}_a\left\{\pi r^2\left(1-{{V(r)}\over{E_\infty}}\right)\right\},
\end{equation}
where ${\rm min}_a\{\}$ denotes the minimum value over all radii $r\ge a$, and it is understood that $\sigma\ge0$.  For the potential profile appropriate to solid \htwo\ (\S3.3), the location of the minimum is independent of particle energy, $E_\infty$. Subsequent averaging over a thermal distribution then leads to exactly the same outcome as equations (19--21), with $\gamma$ being determined by the potential energy evaluated at the location of the minimum in equation (26). 

For our problem there are three relevant cross-sections, with their associated $\gamma$ values: the cross-section for an electron or a proton to reach the surface of the grain, and the cross-section for an electron to penetrate into the solid. For an electron reaching the surface, the minimum in equation (26) occurs just outside the SSE, at $r=a^+$, where $\Phi(a^+)\simeq\Phi_c(a)$, and the Coulomb potential $\Phi_c$ is as per equation (18). Defining ${\cal N}_e\equiv N_e/N_0$ we find the resulting value of $\gamma$ to be
\begin{equation}
\gamma_e^+={\cal N}_p-{\cal N}_e.
\end{equation}
For protons reaching the surface of the grain, the minimum in equation (26) occurs at the surface itself, where the potential is $\Phi(a)=\Phi_c(a)+\delta\Phi_{sse}$. The resulting value of $\gamma$ is thus
\begin{equation}
\gamma_p={\cal N}_e-{\cal N}_p-{\mu\over a}{\cal N}_e.
\end{equation}
Although $\mu\ll a$, the potential step through the SSE is important in circumstances where $N_e\gg |N_p-N_e|$, and the final term in equation (28) cannot be neglected. 

Any positive ion which reaches the grain surface will penetrate the solid, binding into the matrix just below the surface. But electrons reaching the surface will not penetrate the matrix unless they have enough energy to climb into the conduction band of the solid. The corresponding value of $\gamma$ is 
\begin{equation}
\gamma_e={\cal N}_p-{\cal N}_e+{\mu\over a}{\cal N}_e - {{V_0}\over{kT_{ism}}}.
\end{equation}
Strictly, $\gamma_e$ should be the smaller of $\gamma_e^+$ and the value in equation (29).

The evolution of the net surface charge within the solid matrix is then determined by equation (23), with $\gamma_p$ and $\gamma_e$ given by equations (28) and (29), respectively. To calculate how $N_p$ varies, we must simultaneously solve for the evolution of the SSE density, $N_e$. Here we assume that any electron reaching the surface with sufficient energy to penetrate the solid does so, and that all other electrons reaching the surface lose some kinetic energy on impact and become trapped in bound vacuum states. We therefore model the SSE density evolution using
\begin{equation}
{{{\rm d}{\cal N}_e}\over{{\rm d}\tau}}={\cal F}(\gamma_e^+)-{\cal F}(\gamma_e),
\end{equation}
with $\gamma_e^+$ and $\gamma_e$ given by equations (27) and (29), respectively. We can now proceed to solve equations (23) and (30) as coupled differential equations.

\begin{figure}
\includegraphics[width=80mm]{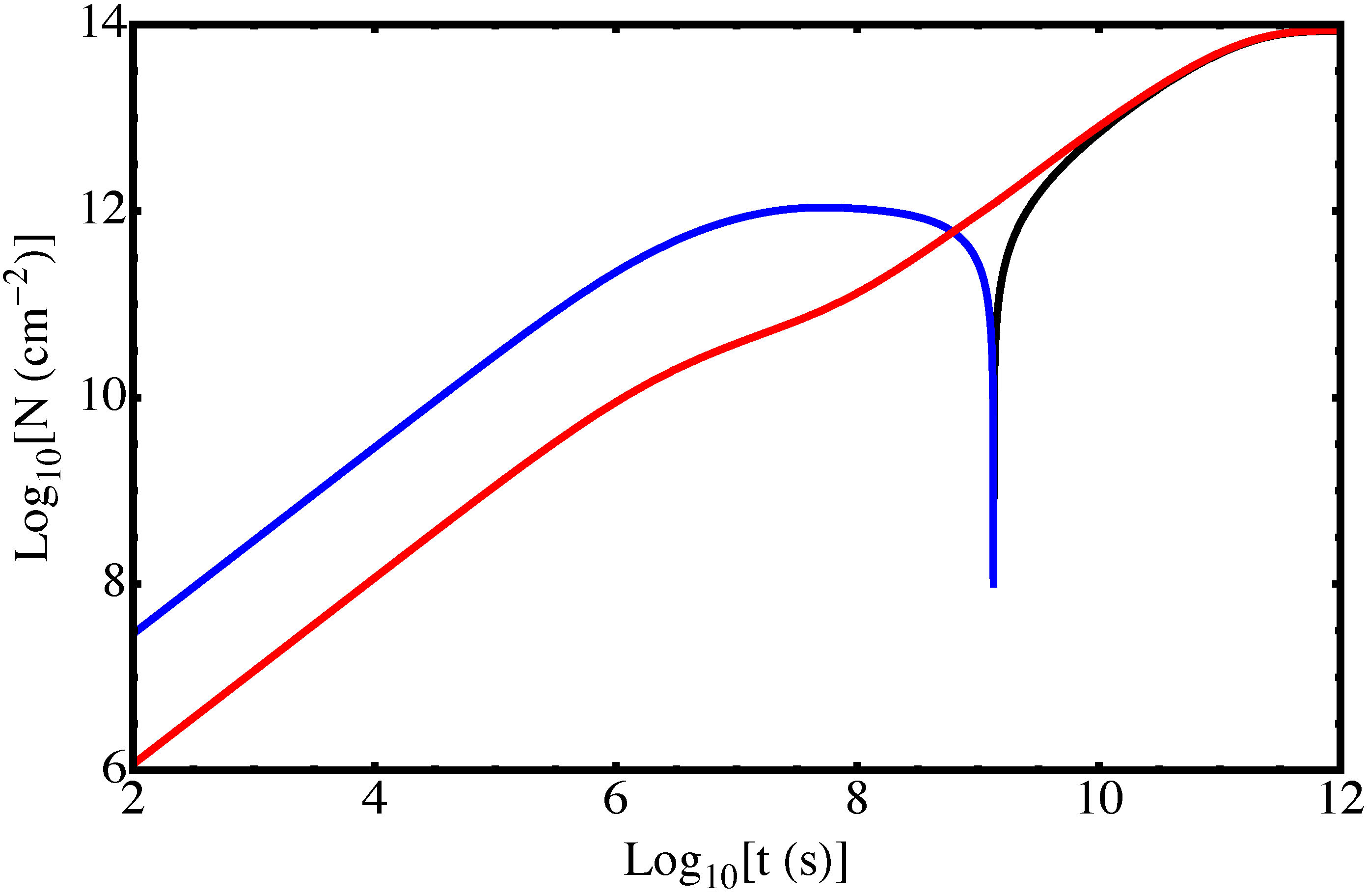}
\includegraphics[width=80mm]{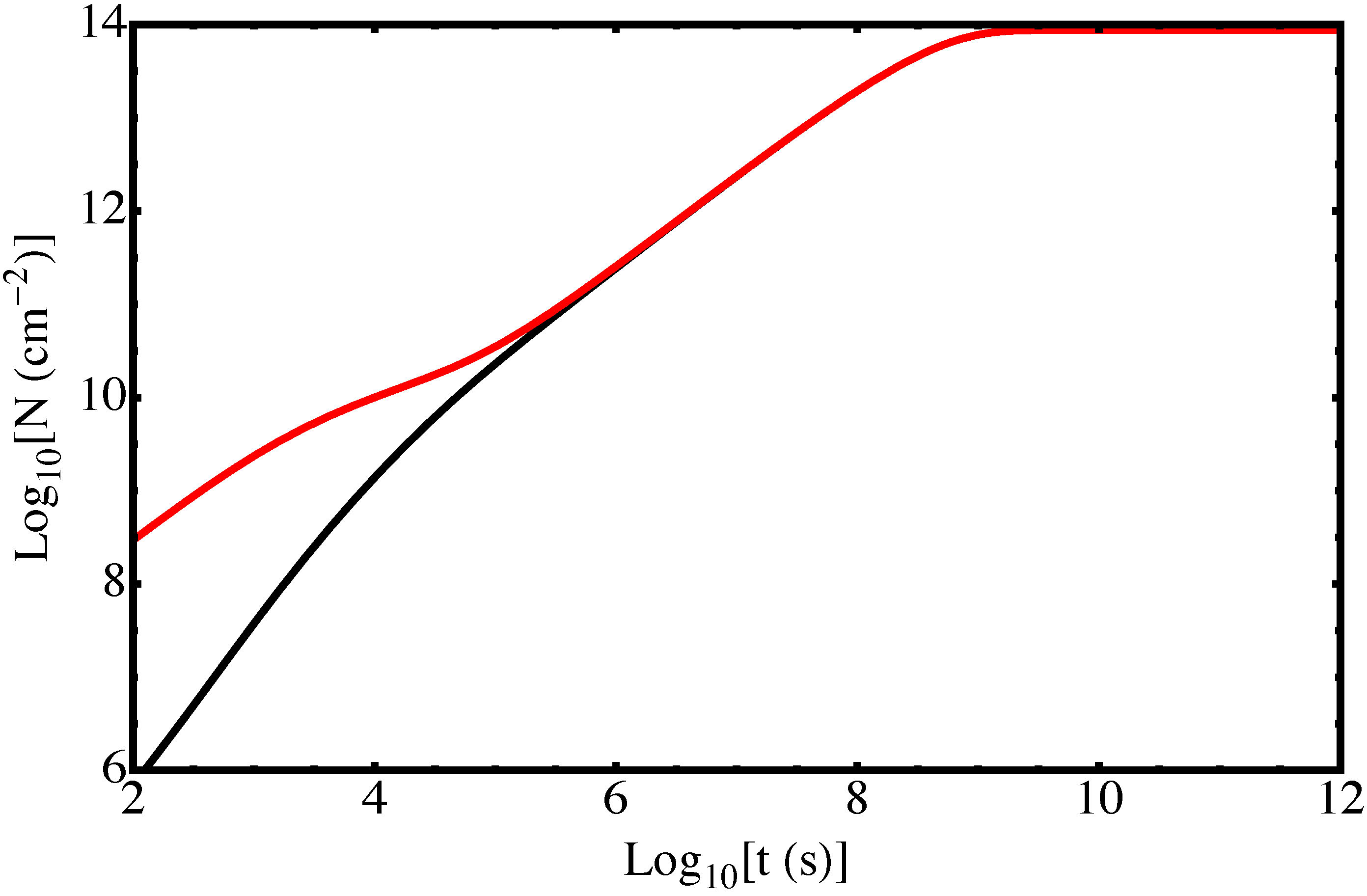}
\includegraphics[width=80mm]{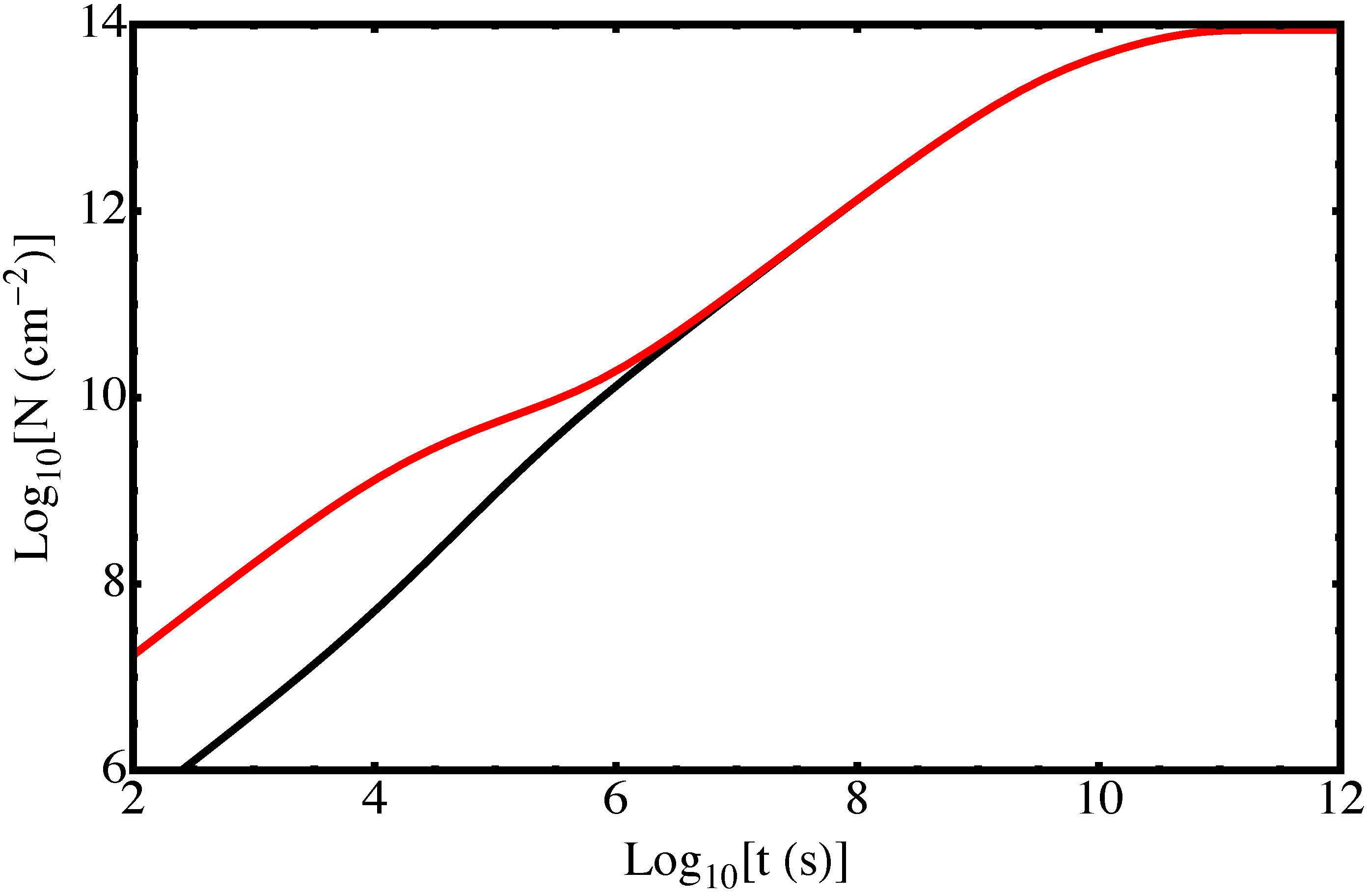}
\includegraphics[width=80mm]{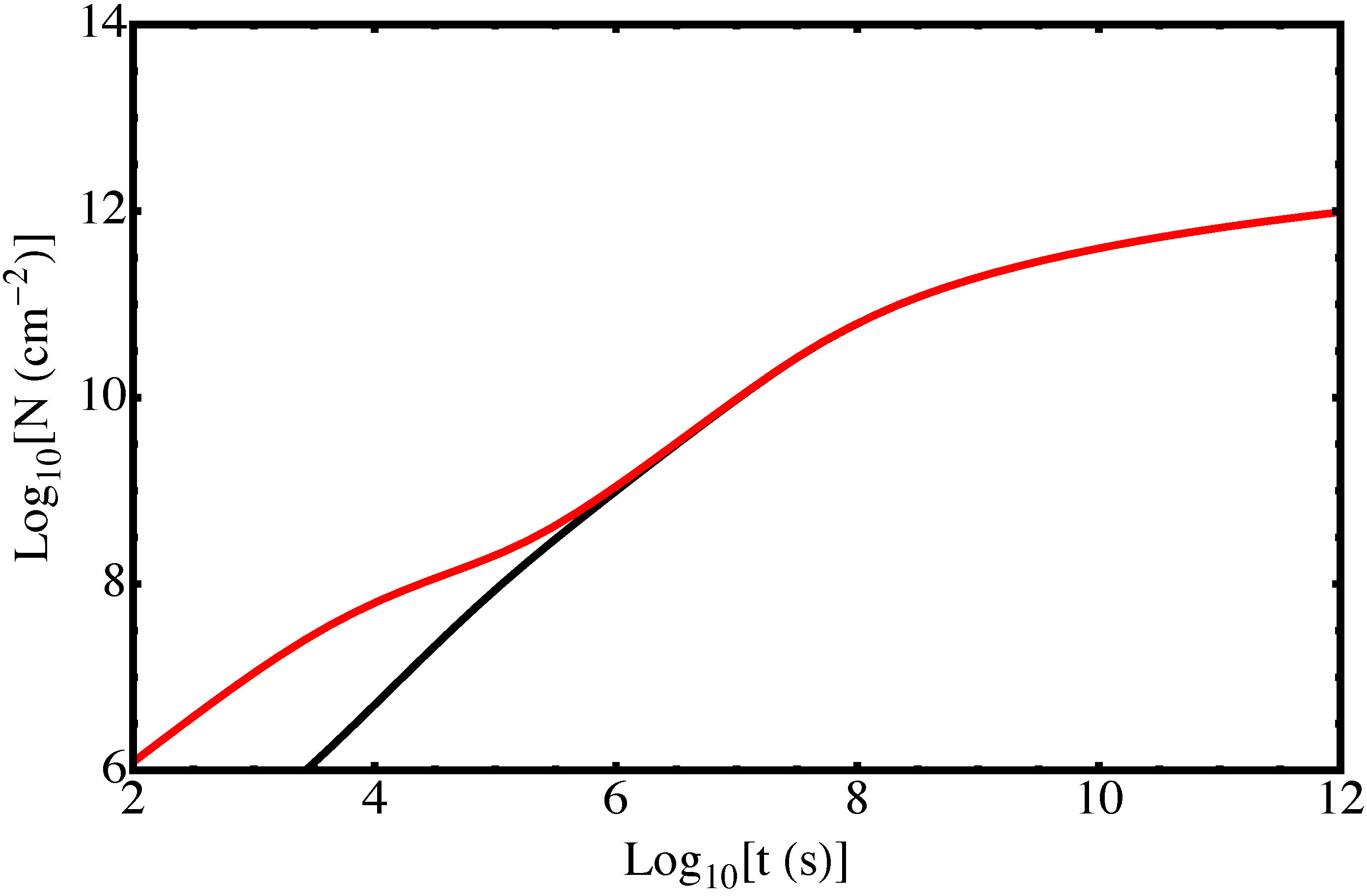}
\caption{The development of charge density on grains of solid \htwo, of radius $a=1\,{\rm\mu m}$, in the case of pure collisional charging. The panels show grains located in different phases of the ISM (top to bottom): HIM; WIM; WNM; and CNM (with characteristics given in table 1). The SSE density is shown in red, whereas the density of charge in the solid matrix is shown in black if it is positive, and blue if it is negative.}
\medskip
\end{figure}

\subsubsection{Solutions for \htwo\ grains}
We have obtained numerical solutions to equations (23) and (30), describing the pure collisional charging of solid \htwo\ grains, in the four different phases of the ISM characterised in table 1. These solutions are shown in figure 3 for a grain radius $a=1\,{\rm\mu m}$. 

To understand how $N_e$ and $N_p$ evolve it is helpful to examine the initial conditions. At $t=0$ we have $N_e=N_p=0$, so $\gamma_e^+=\gamma_p=0$, and $\gamma_e=- {{V_0}/{kT_{ism}}}$. In the high temperature regime ($kT_{ism}\gg V_0$) we therefore expect a small initial growth rate in the SSE density, and a net negative charge accumulating in the matrix itself, with ${\cal N}_p\simeq-\tau$ for $\tau\ll1$. This behaviour is seen in the HIM (top panel of figure 3). Physically it corresponds to the temperature being so high that almost all electrons that reach the surface of the grain have sufficient energy to enter the conduction band and thus penetrate the solid. 

In the low-temperature regime, on the other hand, only a tiny fraction of electrons can penetrate the solid. In this case one expects a density in the SSE which initially grows as ${\cal N}_e\simeq\tau$, and a net positive charge gradually accumulating in the matrix, with ${\cal N}_p\simeq\tau\sqrt{m_e/m_p}$ at early times. This behaviour is seen in figure 3 in the warm and cold phases of the ISM.

The early-time solutions just described are valid until  $\tau\sim1$ ($t\sim t_o$), at which point either ${\cal N}_e$ or ${\cal N}_p$ becomes significant and the corresponding growth rate is moderated. This leads to an approach to overall charge neutrality, with $N_e\simeq N_p$. 

The steady-state conditions implied by equations (23) and (30), at late times, are easily obtained by setting their right-hand-sides to zero. In the case of equation (30) this immediately yields the result
\begin{equation}
N_e\rightarrow N_e(\infty)={{\epsilon_o V_0}\over{e^2 \mu}},
\end{equation}
as $\tau\rightarrow\infty$. This condition can also be written as
\begin{equation}
e\,\delta\Phi_{sse}=V_0,
\end{equation}
i.e. the potential drop through the SSE is equal to the height of the surface barrier. Stated in that way it is easy to see why equation (31) is the limiting charge density: any electron which reaches the top of the SSE will also penetrate into the solid, so the SSE density can grow no further.

At high charge densities the SSE is confined primarily by the electric field of the positive ions in the solid matrix, so it is appropriate to calculate the potential drop $\delta\Phi_{sse}$ using the wavefunctions of \S3.2. Therefore the relevant length scale is $\mu\simeq1.56 s$, and with $s$ as given in equation (15) the limiting charge density $N_e(\infty)$ is determined by noting the near-equality between $N_e(\infty)$ and $N_p(\infty)$. The result is
\begin{equation}
N_e(\infty)=8.6\times10^{13}\;{\rm cm^{-2}},
\end{equation}
corresponding to a limiting surface field
\begin{equation}
{\cal E}\rightarrow{\cal E}_{max}=1.6\times10^{10}\;{\rm V\,m^{-1}}.
\end{equation}
Excepting the case of very small grains in the HIM (see later), these limiting conditions are universal.

It is worth noting that the field estimate in equation (34) is comparable to the limiting fields which may be expected to arise from (i) field ion emission, which Draine and Sutin (1987, citing Muller and Tsong 1969) give as $\sim3\times10^{10}\;{\rm V\,m^{-1}}$ for silicate/graphite grains, and (ii) the spatial gradient in the conduction band energy at the surface, which is $\sim V_0/e l$, where $l$ is the nearest neighbour distance in the solid, with $l^3\rho_*\sim m$, so this evaluates to $\sim10^{10}\;{\rm V\,m^{-1}}$. We do not attempt to model those processes here, so we cannot say whether either of them imposes a limit that is smaller than the one given in equation (34), but the possibility is acknowledged.

Analogous to equation (31), describing the steady-state conditions for ${\cal N}_e$, the asymptotic limit for ${\cal N}_p$ is obtained by requiring the right-hand-side of equation (23) to vanish. In this way we obtain the limit ${\cal N}_p\rightarrow{\cal N}_e-\delta{\cal N}_\infty$, where
\begin{equation}
\left(1+\delta{\cal N}_\infty- {{V_0}\over{kT_{ism}}}\right)\exp[\delta{\cal N}_\infty]=\sqrt{m_p/m_e}.
\end{equation} 
In the high temperature regime, $kT_{ism}\gg V_0$, this condition is approximately the same as equation (25), so that $\delta{\cal N}_\infty\simeq{\cal N}_\infty\simeq2.5$. Whereas in the opposite, low-temperature regime the asymptotic difference in charge densities can be approximated by
\begin{equation}
\delta{\cal N}_\infty \simeq {{V_0}\over{kT_{ism}}}.
\end{equation} 
Although this value is large compared to unity, it is nevertheless a fraction $\mu/a\ll1$ of ${\cal N}_e\simeq{\cal N}_p$ at late times. Consequently we observe near coincidence of $N_e$ and $N_p$ at late times in the CNM (figure 3).

At late times figure 3 shows that the SSE density ($N_e$, red curves) indeed approaches $N_e(\infty)$, in warm and hot media. In the cold, neutral medium, however, both $N_e$ and $N_p$ exhibit very slow growth at late times, and have reached only $\sim0.01\times$ their asymptotic values by $t\simeq3\,$Myr. This regime of slow charging is encountered when the final term in equation (28) reaches a value of order unity, so that the potential drop across the SSE has a significant effect on the proton flux. This regime is never reached in the HIM, and for the warm media it is reached at $N_e\sim N_e(\infty)$ so it has little influence on the result.

For \htwo\ grains of different sizes charging proceeds in a fashion which is qualitatively similar to that of the $a=1\,{\mu m}$ grains shown in figure 3. The main quantitative change with grain size is that $N_e$ and $N_p$ converge at earlier times in the case of larger grains. That is because grain neutralisation is a response to the Coulomb potential becoming significant, and a given Coulomb potential corresponds to $N_e\propto 1/a$. 

In this paper we have restricted attention to grains larger than $0.1\,{\rm\mu m}$. For smaller grains in the HIM the electrostatic potential energy associated with an unshielded charge density $|N_p|\sim N_e(\infty)$ may be small compared to $kT_{ism}$. In this circumstance hydrogen grains charge up in a manner similar to silicate or graphite grains of the same size (i.e. as per figure 2), and the SSE is largely irrelevant. The net charge on such grains is negative, and their surface electric fields are stronger than the value given in equation (34).

The asymptotic limit in equation (31) is different in form, and may be much larger than that obtained in \S4.1 for conventional grain materials. For example, in the WIM a silicate or carbonaceous grain of radius $1\,\mu{\rm m}$ is expected to acquire a (negative) surface charge density $\sim10^{10}\,{\rm cm^{-2}}$, almost four orders of magnitude less than the limit for hydrogen grains.

\section{Sublimation from a charged surface}
Because of the electric field at the surface of the grain, sublimating molecules are polarised, with an associated electrostatic energy $E_{pol}<0$. At high charge densities, where $N_e\simeq N_p$, $|E_{pol}|$ falls rapidly to zero as a molecule is moved outward through the SSE. Sublimation from a polarised surface therefore requires an additional energy $b_{pol}=-E_{pol}$ to be supplied to each sublimating molecule. The sublimation rate and saturated vapour pressure are then given by equations (3) and (2), but with the replacement $b_o\rightarrow b_o+b_{pol}$. We now evaluate $b_{pol}$.

\subsection{Electrostatic binding}
The molecular polarisability, $\alpha$, of \htwo\ was calculated to high precision by Ko\l os and Wolniewicz (1967). For the rovibrational ground state they obtain (in atomic units)\footnote{To convert to SI units (${\rm F\,m^2}$), multiply by $1.6488\times10^{-41}$.}
\begin{equation}
\alpha_\parallel=6.7632,\qquad\alpha_\perp=4.7393,
\end{equation}
parallel and perpendicular, respectively, to the molecular axis. If there is no preferred orientation of the molecules one works with the spherical average of these values, $\langle\alpha\rangle=(\alpha_\parallel+2\alpha_\perp)/3$. But in our case the electric fields, ${\cal E}$, are sufficiently strong that the magnitude of the molecular polarisation energy
\begin{equation}
|E_{pol}|={1\over2}\left\{\alpha_\parallel{\cal E}_\parallel^2+\alpha_\perp{\cal E}_\perp^2\right\},
\end{equation}
is large compared to the typical thermal energy in the solid, and the surface molecules will preferentially orient themselves\footnote{Alignment also alters the van~der~Waals forces between neighbouring molecules, but in this paper we neglect that effect.}  so that ${\cal E}_\perp^2\ll{\cal E}_\parallel^2\simeq{\cal E}^2$.

In \S4 we estimated the electric field at the limiting charge density to be ${\cal E}_{max}\simeq1.6\times10^{10}\,{\rm V\,m^{-1}}$. From this we can now evaluate the electrostatic contribution to the heat of sublimation, it is
\begin{equation}
b_{pol}=-E_{pol}\simeq{1\over2}\alpha_\parallel{\cal E}_{max}^2\simeq84\;{\rm meV}.
\end{equation}
Hence $b_{pol}/k\simeq980\;$K. This is an order of magnitude larger than $b_o$, demonstrating that grain charging has a profound effect on the thermodynamics of the surface.
\begin{figure}
\includegraphics[width=85mm]{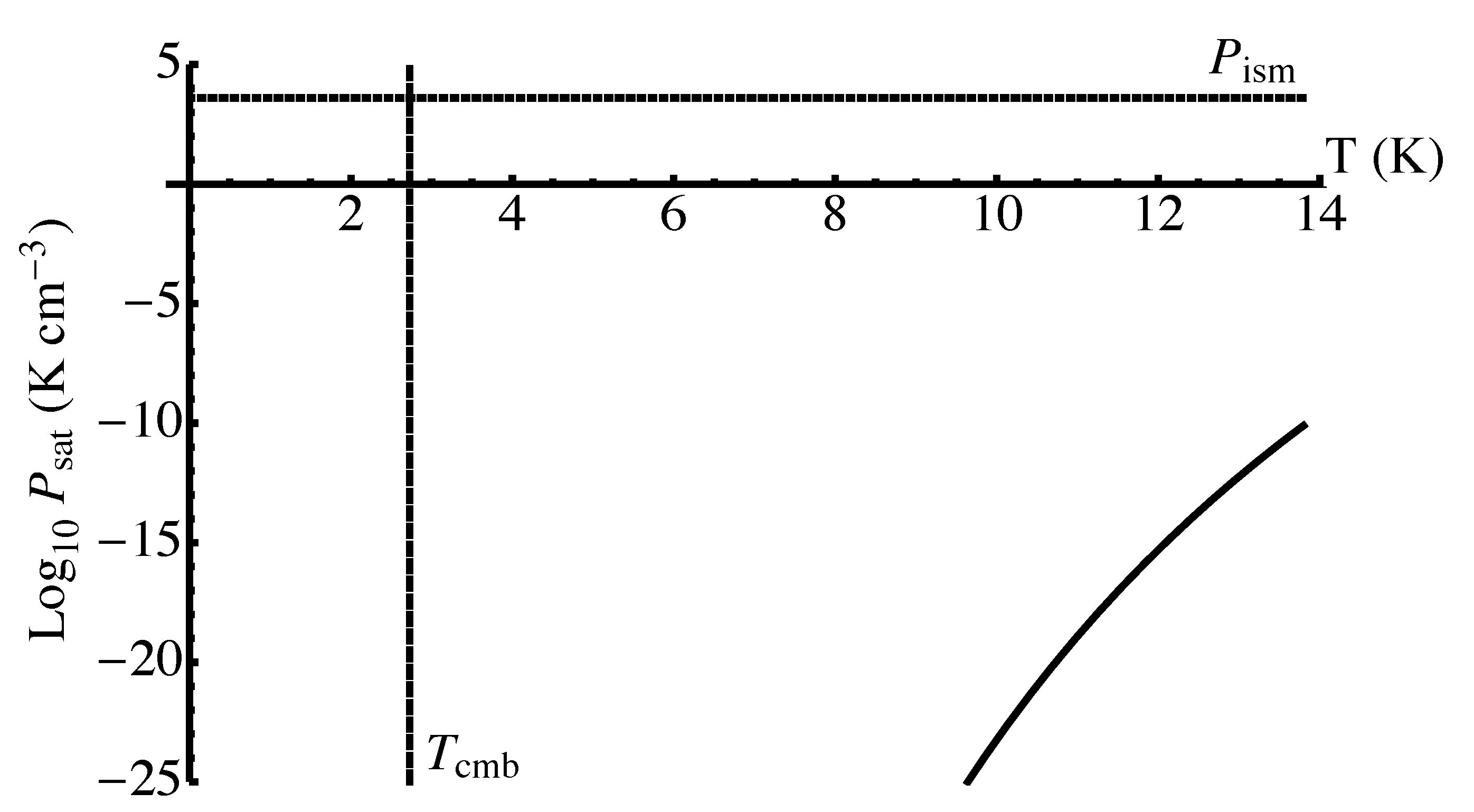}
\vskip0.truecm
\caption{As figure1, but for the case of a fully charged \htwo\ surface. Note that $P_{sat}\ll P_{ism}$ at all temperatures below the triple-point, suggesting that grains of \htwo\ may be able to live indefinitely in the diffuse ISM. Our estimated radiative equilibrium temperature (i.e. neglecting sublimation cooling) for a micron-sized grain of \htwo\  is $7.4\,$K (\S2.2).}
\medskip
\end{figure}

Figure 4 shows the influence of $b_{pol}/k=980\,$K on the saturated vapour pressure of \htwo. Comparing to figure 1 we see that saturation for the charged surface occurs at pressures which are many orders of magnitude below that for the uncharged case. Indeed figure 4 shows that our estimate of the saturated vapour pressure for a fully charged surface lies many orders of magnitude below $P_{ism}$ at all temperatures below the triple-point. 

We caution that our estimate of ${\cal E}_{max}$ obtained in \S4 is a crude one, and $b_{pol}$ varies as the square of ${\cal E}_{max}$ so the particular value we have obtained should not be given undue emphasis. Nevertheless our calculation serves to demonstrate that grain charging plays a critical role in determining the sublimation rate of interstellar solid hydrogen, and opens up the possibility of \htwo\ grains surviving indefinitely in the diffuse ISM.

\section{Discussion}
Due to a lack of information on yields, we neglected photoelectric charging. We now estimate how small the photoelectric yield would have to be in order  for that approximation to be good. Because the band-gap of solid \htwo\ is greater than the ionisation energy of atomic hydrogen, the number-density of photons in the ISRF which can excite electrons into the conduction band is very small. Therefore the contribution of such photons to the charging rate is limited even if their photoelectron yield is high. Far-UV photons are much more numerous up to energies of one Rydberg, and some of those photons may yield photoelectrons because the excitonic states which they induce in the solid lie above the vacuum free-electron energy --- excitons could decay if they diffuse to the surface of the \htwo\ crystal. Using the far-UV ISRF energy density given in \S2.2, we estimate the photoelectric emission rate of an uncharged solid hydrogen surface to be $3.7\times10^6\eta\,{\rm cm^{-2}\,s^{-1}}$, where $\eta$ is the yield. Comparing this with the lowest collisional rate (i.e. that for the CNM; table 1), we see that we need $\eta\ll3\times10^{-3}$ for photoelectric emission to be negligible. We also note, from figure 3, that after $t\sim10^8\,$s the collisional charging rate in the CNM declines substantially, and in that regime of the charging process the yield would have to be much smaller still in order for photoemission to be negligible. Laboratory investigation of the far-UV photoelectric yield of solid \htwo\ would be valuable. 

Figure 3 shows that pure collisional charging in the CNM does not permit a micron-sized grain to reach the fully charged state within the sublimation time-scale ($6\times10^9\,$s, see \S2.3). That does not mean that such a grain will vanish. Our charging calculations do not include the effect of sublimation of \htwo, which leaves the charged species in place but decreases the surface area of the grain. Thus even in the case of a hydrogen grain in the CNM with zero photoemission, the particle will not disappear but will shrink until the charge density is high enough to suppress the sublimation rate.

Our finding that charging has a strong influence on the viability of hydrogen grains motivates improvements in the description of charging. Notable deficiencies of the treatment we have given include our use of a single-particle description of the density distribution, and approximating the band-structure of the solid by an infinite potential wall. These deficiencies could be addressed by the application of Density Functional Theory. Spatial structure in the plane of the surface would also be worth investigating, as the confining electric field provided by the sub-surface ions is only approximately uniform. The non-uniformities are interesting because they imply spatial variations in the heat of sublimation of \htwo, which we have not allowed for in our model. They are also interesting in respect of the possibility of confinement of individual electrons in the vicinity of individual ions --- forming a new, and highly unusual sort of ``atom''. Finally we note that the photo-detachment of surface-state electrons and their tunneling recombination with sub-surface ions are processes whose rates should be quantified. In this paper we have implicitly assumed that these rates are negligible in comparison with the collisional charging rates.

It would be valuable to have a better estimate of the radiative equilibrium temperature of a hydrogen grain. A key area of uncertainty in that calculation lies with the coupling of the grain to microwave radiation. Our treatment (\S2.2) was appropriate to a homogeneous grain of insulating material, with a low level of absorption assumed (i.e. $\lambda_2=1\,{\rm\mu m}$, yielding an absorption efficiency $\simeq\,$5\% that of a silicate grain of the same size). The surface charge structure which develops on solid \htwo\ will enhance the microwave coupling of hydrogen grains in two ways. First, we expect strong inhomogeneities in the electric field within the layer of ions bound inside the solid. Consequently any lattice vibrations in that layer will induce large, oscillating electric dipoles and will thus radiate more efficiently than an uncharged grain. Secondly, the surface-state electrons are expected to couple strongly to the microwave radiation field and will contribute to absorption as a result of excitation into higher quantum states.

Sublimation is not the only process which might limit the lifetime of interstellar grains. Sputtering, which we have completely neglected, could play an important role. There have been some laboratory studies of the sputtering of solid \htwo\ by energetic particles (Pedrys et al 1997; Schou 2002), which could provide a basis for estimating its effect in an interstellar context. However, the experimental investigations have presumably been undertaken under conditions of low surface charge density. It is possible that sputtering could be strongly suppressed under conditions of maximal charge, so further laboratory studies may be needed.

\subsection{Paths to discovery of solid hydrogen grains}
The surface-state electrons described in \S3 take the form of a two-dimensional electron gas. As such they imply a metallic character to a charged \htwo\ surface, even though the bulk material is itself an insulator, and a strong interaction with radio-waves is expected. The volume fraction occupied by dust in most of the ISM is small enough (Purcell 1969) that we don't expect dust to have much influence on radio-wave propagation in typical regions of interstellar space. However, one can imagine atypical regions containing neutral helium gas (which is difficult to detect), plus a substantial volume fraction of hydrogen grains, and in such regions \htwo\ particulates could make a significant contribution to the radio-wave refractive index.  Moreover they would do so without any contribution to the fluid pressure. Thus hydrogen grains may offer an explanation for circumstances where heavy scattering arises from regions of small spatial extent. Three such phenomena are known: the Extreme Scattering Events (Fiedler et al 1987, 1994); the intra-hour variations of compact radio quasars (Kedziora-Chudczer et al 1997; Dennett-Thorpe and de~Bruyn 2000; Bignall et al 2003; Lovell et al 2008); and the pulsar parabolic arcs (Stinebring et al 2001; Cordes et al 2006). Indeed hydrogen dust could give rise to apparent source flux variations across a broad range from radio to X-ray, and we note that panchromatic variability is observed in some quasars and BL Lac objects (e.g. Wagner and Witzel 1995).

We have already noted the possibility of forming a new type of  ``atom'' on the surface of solid \htwo, with specific SSE electrons bound to specific, sub-surface positive ions. These electrons would be localised in all three dimensions and would thus have a discrete spectrum of eigenstates, rather than the continua associated with a two-dimensional electron gas (equations 10 and 14). We make no attempt here to calculate the eigenstates of any fully-localised electrons. We simply draw attention to the implication that there should be a set of spectral lines which are characteristic of the charged surface of any interstellar \htwo\ grains. Given that electrons are separated from positive ions by several \AA ngstr\"oms (i.e. at least one crystal plane), the characteristic energies should be a fraction of those for the hydrogen atom. But they cannot be arbitrarily small, because at large electron-ion separations the field of other ions becomes relevant and there will be no unique pairing of a specific electron with a specific ion. Therefore the spectroscopic signature of these ``atoms'' will be confined to a spectral band whose upper boundary lies at photon energies of a few eV. The lines should not be expected to be sharp because there are perturbations from adjacent ions and from other SSE electrons; tunneling phenomena could also make a significant contribution to linewidth. These aspects are reminiscent of the Diffuse Interstellar Bands (Herbig 1995; Sarre 2006).

In contrast to models based on silicates and graphites (e.g. Zubko, Dwek and Arendt 2004), there is no difficulty in explaining the requisite volume fraction of interstellar dust (Purcell 1969) if it is made of hydrogen. Furthermore the size spectrum required for hydrogen grains to fit the extinction data could be very different to that needed for silicate/graphite grains (Mathis, Rumpl and Nordsieck 1977).  For example, hydrogen dust in the ISM might be much bigger than the predominantly $\sim0.1\,{\rm\mu m}$ grains which are required by silicate/graphite models, and might contain much more mass in total. If so, hydrogen grains may be able to explain spacecraft and radar observations of interstellar dust particles streaming through the Solar System (Gr\"un et al 1993; Taylor, Baggaley and Steel 1996; Landgraf et al 2000; Meisel, Janches and Matthews 2002; Mann 2010; Musci et al 2012). These data indicate large numbers of large particles ($\ga1\,{\rm\mu m}$) amongst the interstellar grains, and are in tension with silicate/graphite models both in respect of the abundances of the constituent elements and the shape of the interstellar extinction curve (Draine 2009).

\section{Conclusions}
All types of interstellar dust grains are expected to acquire charge, but the ``double layer'' charge configuration of solid \htwo\ is unique amongst candidate grain materials. Electrons bound in vacuum states above the surface shield the positive charge of ions buried in the solid, so that grains can maintain overall neutrality and charging can continue to high surface-densities of both species. Interstellar grains of solid \htwo\ are thus expected to develop large electric fields within a few \AA ngstr\"oms of their surfaces. Polarisation of \htwo\ molecules enhances their binding to the solid, and for a fully charged surface the heat of sublimation is greatly increased. Charging must therefore be taken into account when determining the sublimation rate and lifetime of  solid \htwo\ in astrophysical environments. It is a mistake to exclude hydrogen grains on the basis of the volatility of the pure solid. 

Electrons bound to the grain surface may be localised in only one dimension, or in all three. The former case yields a two-dimensional free-electron gas, giving the grain surface a metallic character. Electrons which are localised in all three dimensions possess a discrete spectrum of eigenstates, implying that \htwo\ grains may be identified spectroscopically.

\section*{Acknowledgments}
I thank Sterl Phinney for making available a copy of his unpublished manuscript ``Cosmic Snowballs'' (1985), Stefan Bromley for an illuminating conversation on the modelling of the SSE, Artem Tuntsov for some good discussions on all the relevant physics, and Oxford Astrophysics for hospitality.

\label{lastpage}
\end{document}